\def\K{{\rm\thinspace K}}

\def\Msun{\hbox{$\rm\thinspace M_{\odot}$}}

 \def\yr{{\rm\thinspace yr}}

\def\Msunpyr{\hbox{$\Msun\yr^{-1}\,$}}

\documentclass[multphys,vecphys]{svmult}

\usepackage{makeidx}         
\usepackage{graphicx}        
\usepackage{multicol}        
\usepackage[bottom]{footmisc}

\makeindex             

\begin{document}

\title*{Heating and Cooling in the Perseus Cluster Core}
\titlerunning{Perseus cluster}
\author{A.C. Fabian\inst{1}\and J.S. Sanders\inst{2}}
\institute{Institute of Astronomy, University of Cambridge, Madingley
  Road, Cambridge, CB3 0HA, UK.
\texttt{acf@ast.cam.ac.uk}
\and  \texttt{jss@ast.cam.ac.uk}}
%
%
\maketitle

\section{Introduction} 

It is well known that the radiative cooling time of the hot X-ray
emitting gas in the cores of most clusters of galaxies is less than
$10^{10}\yr$ (Fig.~1). In many clusters the gas temperature also drops
towards the centre. If we draw a causal connection between these two
properties then we infer the presence of a cooling flow onto the
central galaxy (e.g.\cite{ Fab94}). High spectral resolution XMM-Newton
data \cite{Pet01}\cite{Tam01} and high spatial
resolution Chandra data, e.g. \cite{All01}, show however a lack of X-ray emitting
gas below about one third of the cluster virial temperature.  The
explanation is that some form of heating balances cooling. The
smoothness and similarity of the cooling time profiles and the
flatness of the required heating profiles all indicate that we must
seek a relatively gentle, quasi-continuous (on timescales $<10^8\yr$),
distributed heat source. The likely such source is the central black
hole and its powerful jets which create bubble-like cavities in the
inner hot gas (\cite{Chu00}, see \cite{Pet06} for a review).

\begin{figure}
  \centering
  \includegraphics[width=0.48\columnwidth]{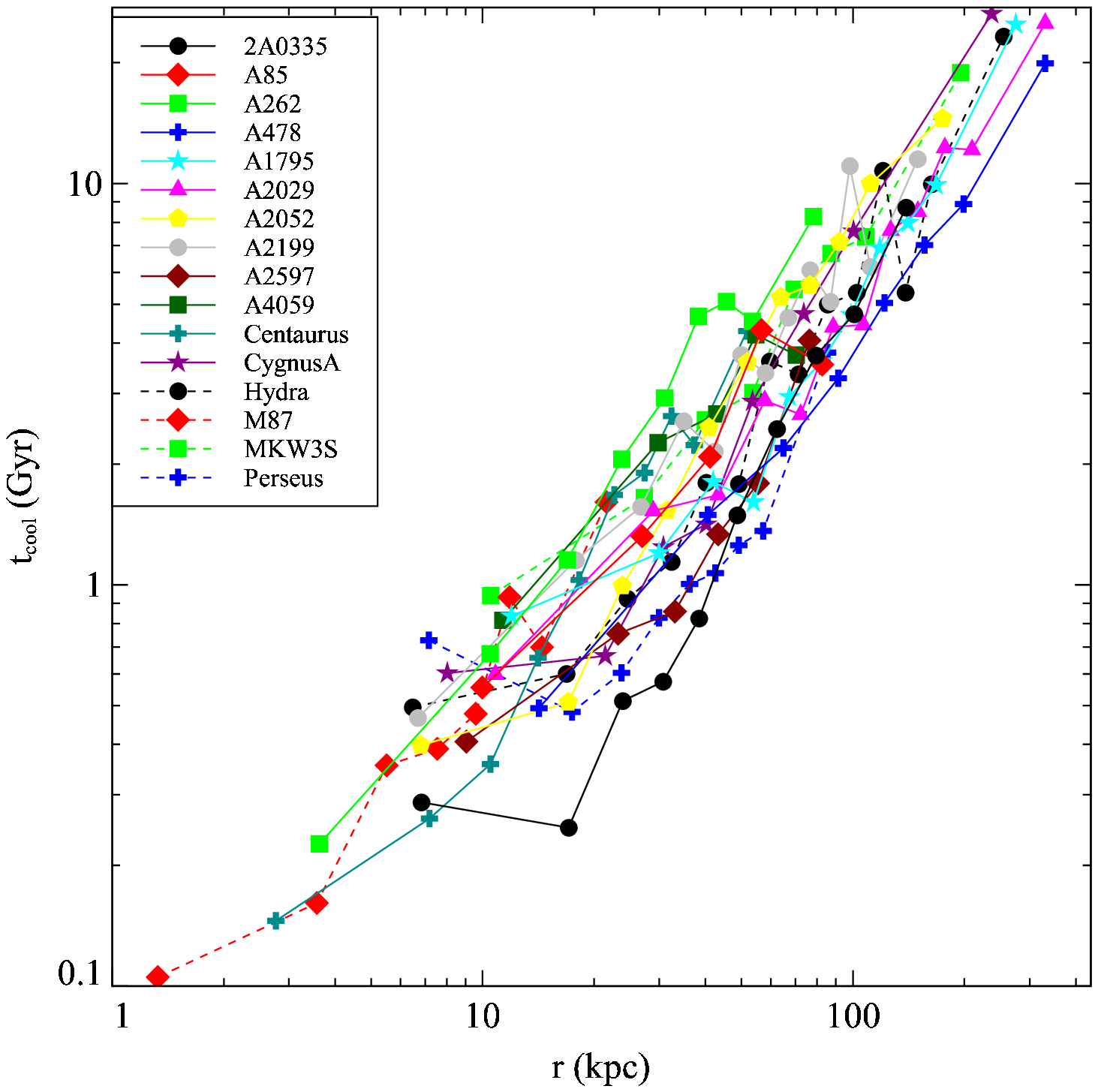}
  \includegraphics[width=0.48\columnwidth]{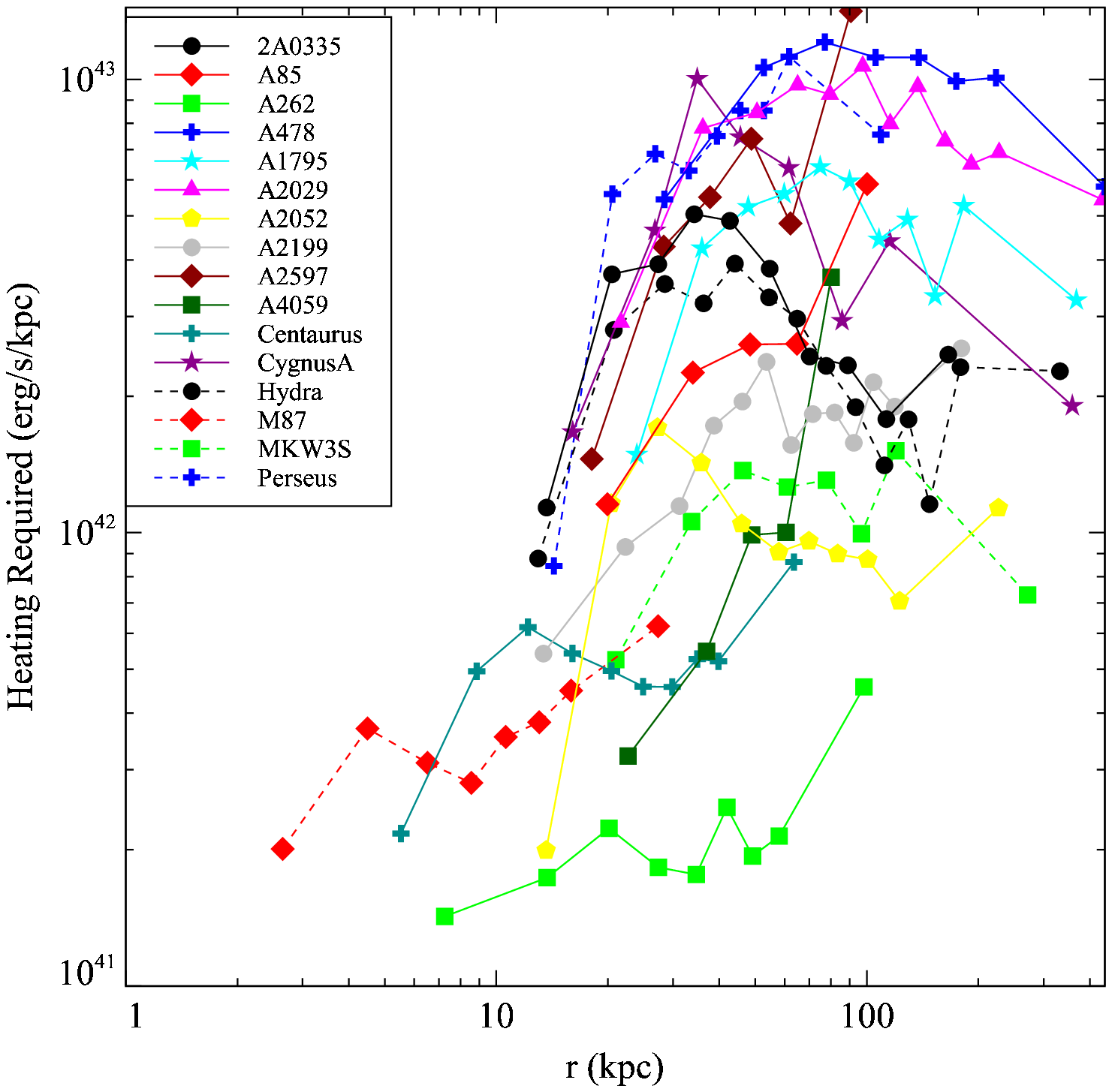}
  \caption{Heating and cooling properties of the intracluster gas in the
    cores of 16 X-ray bright clusters \cite{Dun06}. (Left)
    radiative cooling times and (Right) heating rate required
    per kpc to balance cooling.   }
\end{figure}
\begin{figure}
  \centering
  \includegraphics[width=0.65\columnwidth]{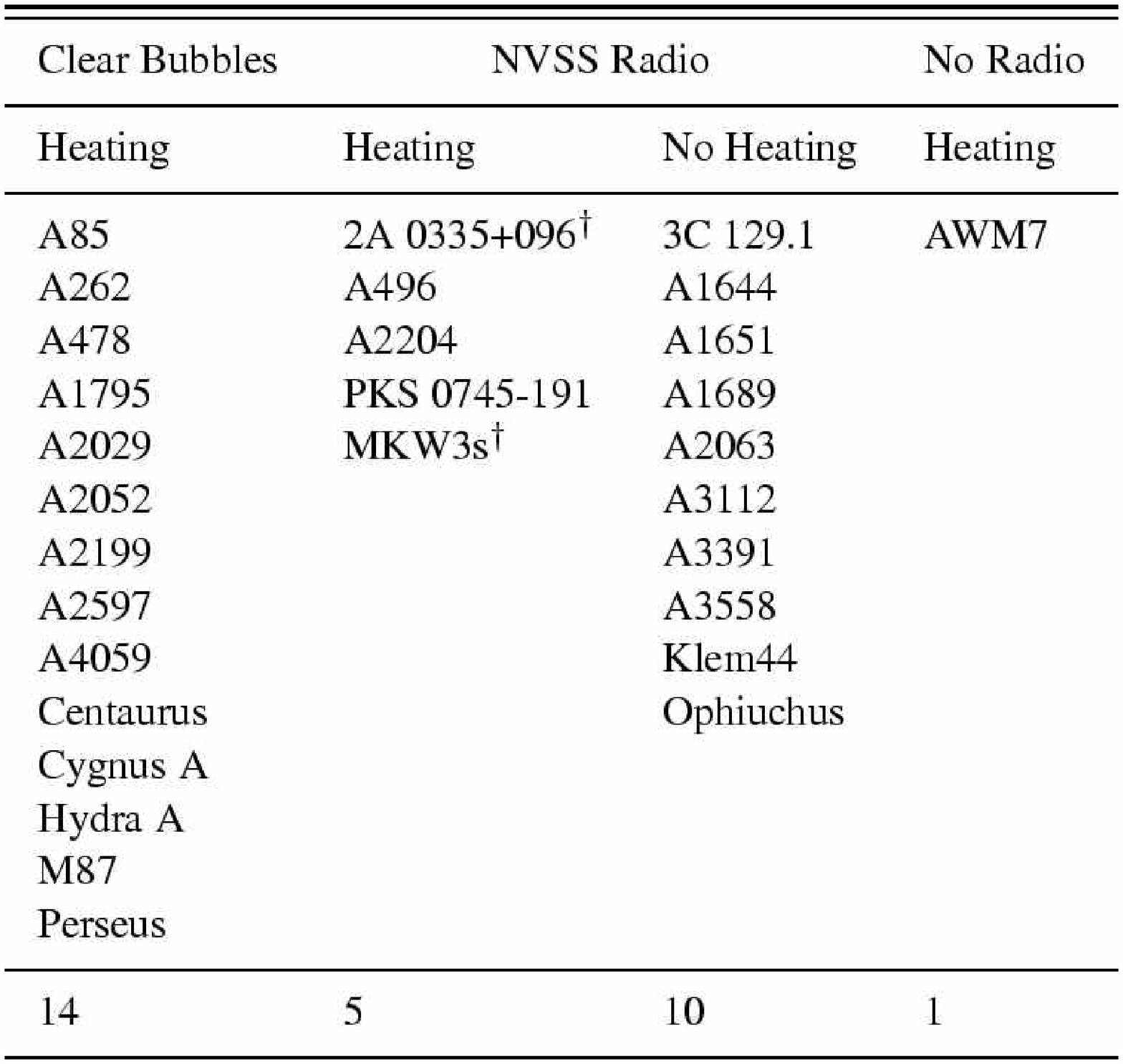}
  \caption{Status of the cores in clusters from the Brightest 55
    sample with Chandra data \cite{Dun06}. Objects in the
    column on the left show clear bubbles, those in the 2 centre
    columns have a central radio source and the one on the right
has no reported radio source at the centre. Heating is defined as
required if the central cooling time is less than 3~Gyr and the
central temperature drop exceeds a factor of two.  
}
\end{figure}
\begin{figure}
  \centering
  \includegraphics[width=0.7\columnwidth]{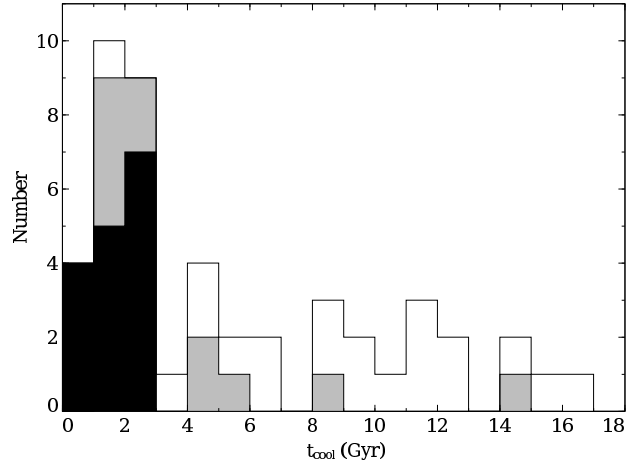}
  \caption{Frequency histogram of the clusters tabulated in
    Fig.~2. Black shading indicates that clear bubbles are seen; grey
    means that there is a plausible radio source at the centre \cite{Dun06}.}
\end{figure}

\begin{figure}
  \centering
  \includegraphics[angle=-90,width=0.7\columnwidth]{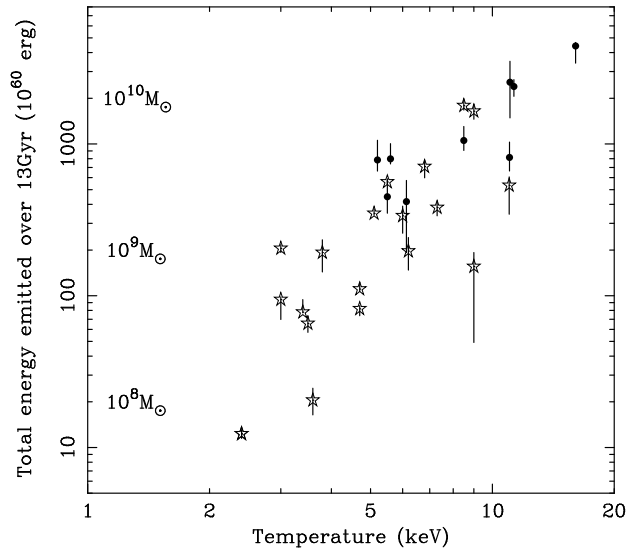}
  \caption{An approximate estimate of the total energy emitted within
    the cooling radius over 13~Gyr plotted versus cluster temperature
    \cite{Fab02a}. The accreted black hole mass, assuming a
    radiative efficiency of 0.1, is also shown. The results reduce
    roughly as (time)$^2$, so drop by 7 if a cluster age of only 5~Gyr
    is assumed.The accreted masses are then still considerable (up to
    $3\times 10^9\Msun$).  }
\end{figure}

We briefly review the {\it general} heating and cooling statistics in
an X-ray bright sample of cluster before we discuss the {\it detailed}
situation in the Perseus cluster, the X-ray brightest cluster in the
Sky. The Chandra count rate from the Perseus cluster is twice that of
any other cool core cluster (the Virgo and Centaurus clusters come
next) and the total exposure, of almost 1~Ms, is almost twice as long
(compare with 500~ks for Virgo and 200~ks for Centaurus). In many ways
therefore the Perseus cluster data are the best, and possibly unique,
for determining the details of the energy flow in a cluster core.

\section{General conclusions from the Brightest Clusters}

Here we summarize the general situation in low redshift clusters using
the Chandra analysis of the Brightest 55 sample from \cite{Dun06}.
The results, where relevant, are in agreement with the earlier ROSAT
analysis of \cite{Per98}. Fig.~2 lists 30 sources of which only
one (AWM7) does not have a radio source associated with the central
galaxy. 14 of them have clear bubbles and another 5 have possible
bubbles or interaction between the radio source and the hot gas. Only
10 out of the 30 do not need any heating (i.e. either the central
cooling time exceeds 3~Gyr, see Fig.~2, and/or any central temperature
drop is less than a factor of two).

The common occurrence of bubbles in the clusters which require heating
means that the duty cycle of bubbling is between 75 and 90 per cent --
the jets are 'on' most of the time. Such a mechanism satisfies the
quasi-continuous requirement.

The steep abundance gradient observed in many clusters requires that
the the heating mechanism be relatively gentle and not produce much
turbulence \cite{Reb05}\cite{Gra06}. The relatively smooth and similar
cooling time profiles support this. 

The energy requirements are considerable (Fig.~4 from \cite{Fab02a}).
The $Pd V$ work done by the bubbles on the surrounding gas is
nevertheless of the right order of magnitude
\cite{Bir04}\cite{Raf06}\cite{Dun06}. It appears inescapable that the
central radio source is pumping energy into the intracluster gas at
roughly the required level. What is not clear from a general study of
clusters, albeit the brighter ones, is how that energy is distributed
on the right spatial scale to balance radiative cooling.  Gas is not
piling up at any particular radius or temperature. An even more
challenging issue is how the feedback required for a tight heating /
cooling balance operates.
 
Highly energetic events, such as inferred for Hydra A \cite{Nul05},
whilst spectacular, are not typical of the norm. They also dump most
of their energy well beyond the coooling region. 

\section{The Perseus Cluster}

\begin{figure}
  \centering
  \includegraphics[width=0.98\columnwidth]{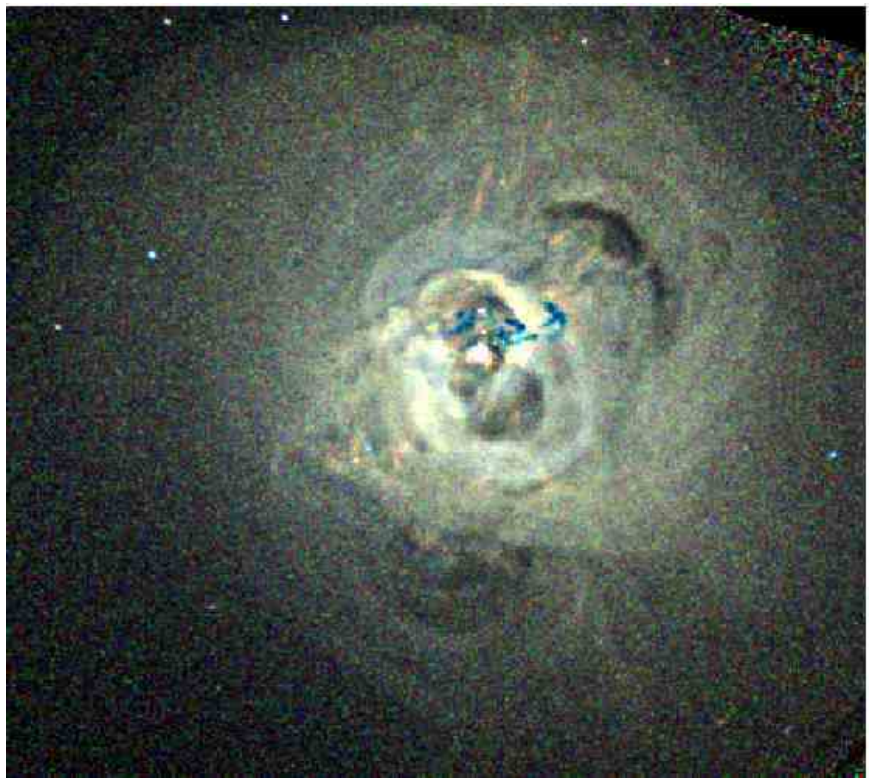}
  \caption{Full 900 ks Chandra image of the centre of the Perseus
    cluster with red, green and blue representing soft, medium and
    hard X-rays within the Chandra band \cite{Fab06}.}
\end{figure}

The Perseus cluster, A\,426, is the X-ray brightest cluster in the Sky.
The X-ray 
emission is  sharply peaked on the cluster
core, centred on the cD galaxy NGC\,1275. Jets from the nucleus of
that galaxy have inflated bubbles to the immediate N and S, displacing
the ICM \cite{Boh93}\cite{Fab00}. Ghost bubbles
devoid of radio-emitting electrons, presumably from past activity, are
seen to the NW and S. The radiative cooling time of the gas in the
inner few tens of kpc is 2--3 hundred Myr, leading to a cooling flow of a
few $100\Msunpyr$ if there is no balancing heat input. Energy from the
bubbles or the bubble inflation process is a likely source of heat but
the energy transport and dissipation mechanisms have been uncertain. 

\begin{figure}
 \centering
  \includegraphics[width=0.47\columnwidth]{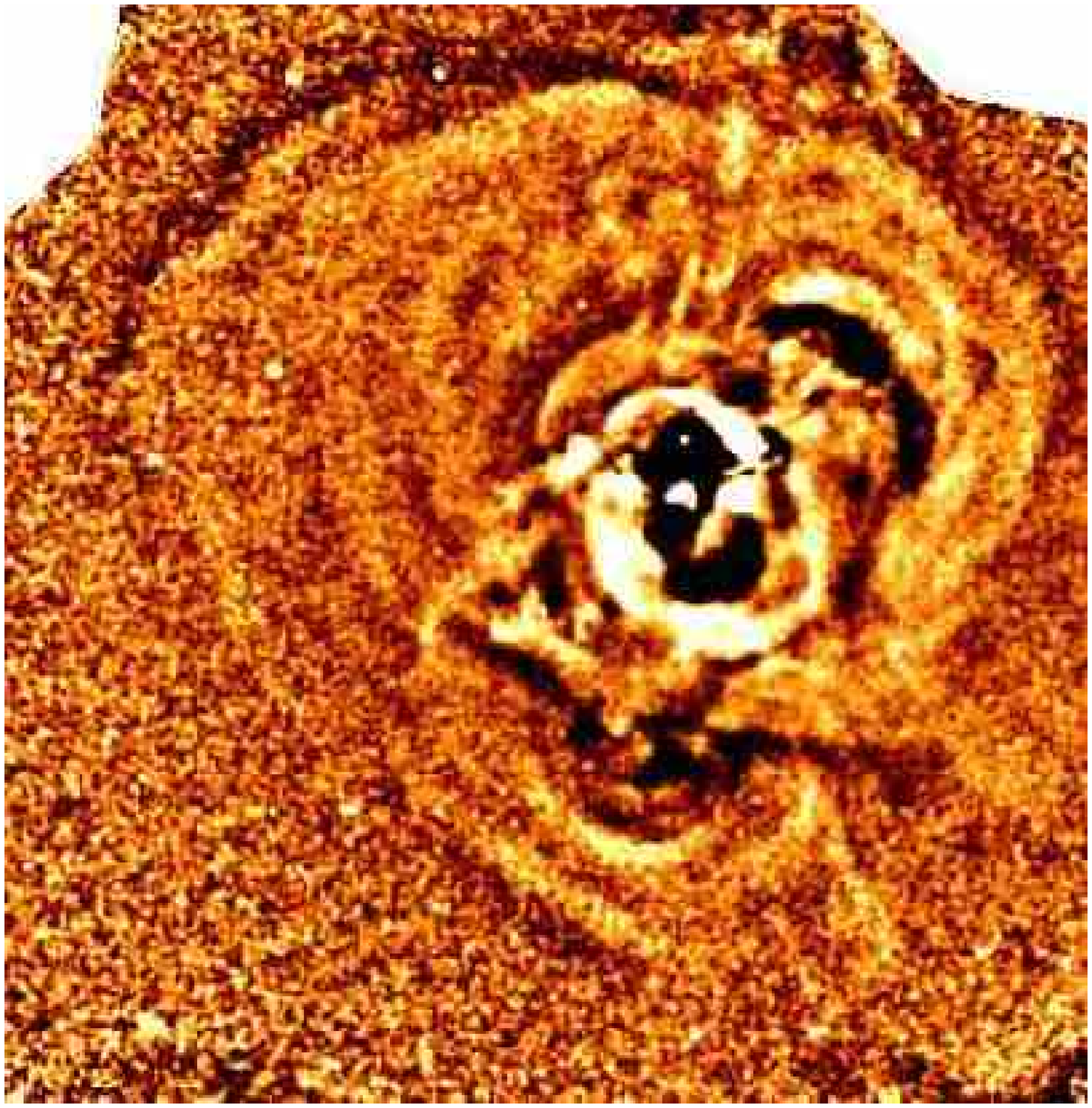}
 \includegraphics[width=0.47\columnwidth]{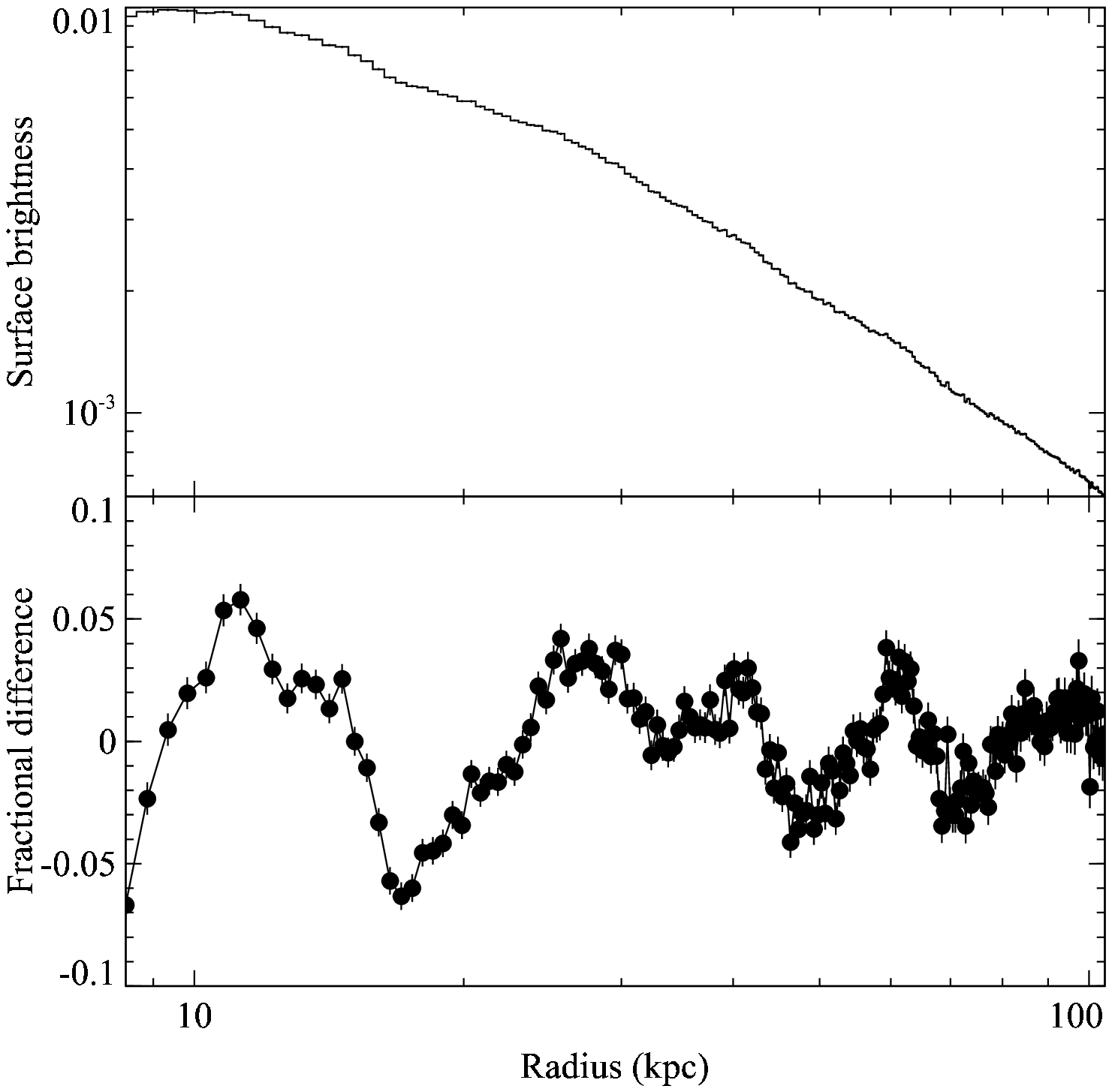}
  \caption{Left: unsharp-mask image showing the ripples, Right: profile
    of the raw data to the E with deviations from a simple beta-model
    fit to that profile.}
\end{figure}  

\begin{figure}
  \centering
  \includegraphics[width=0.32\columnwidth]{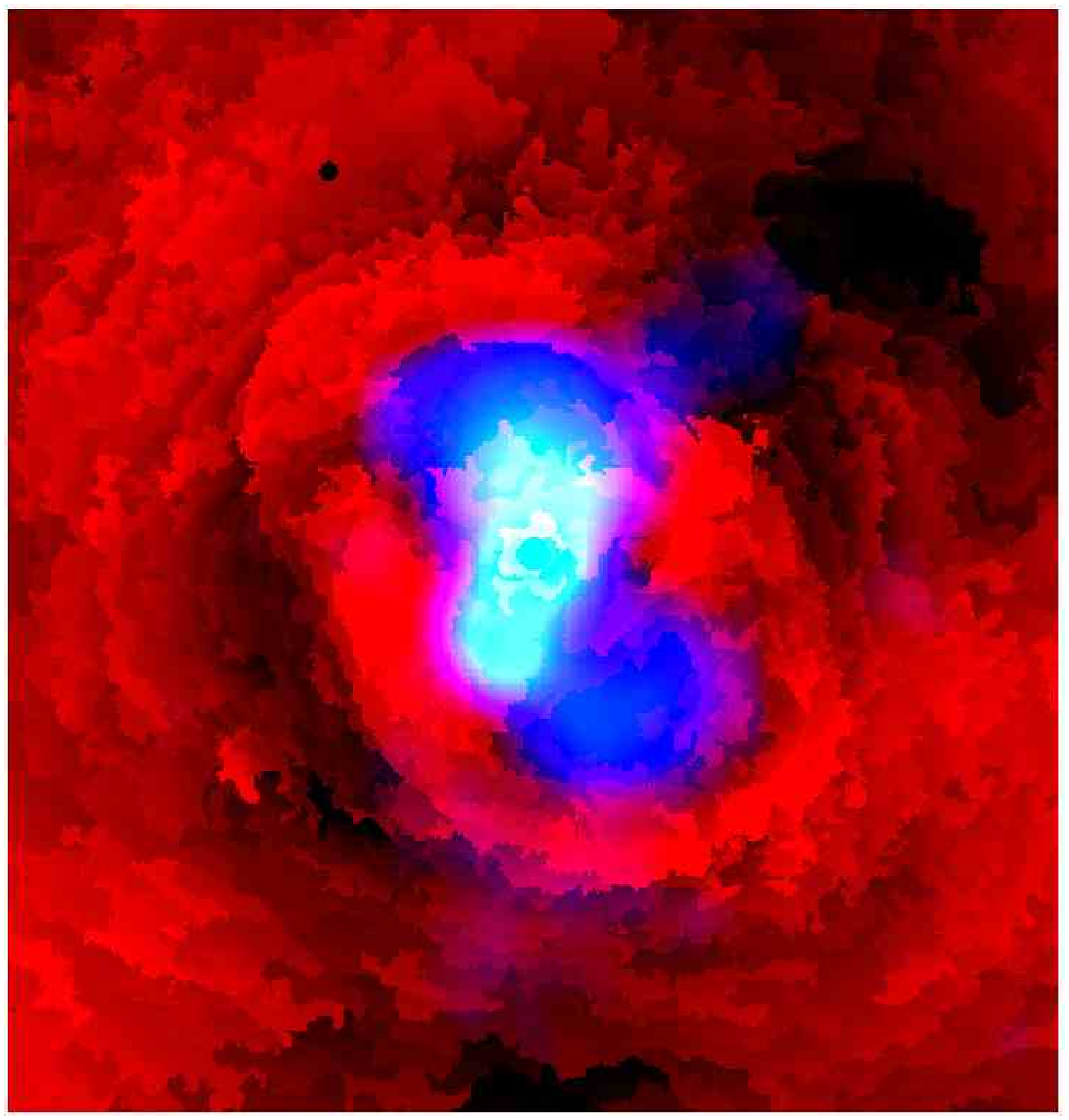}
  \includegraphics[width=0.32\columnwidth]{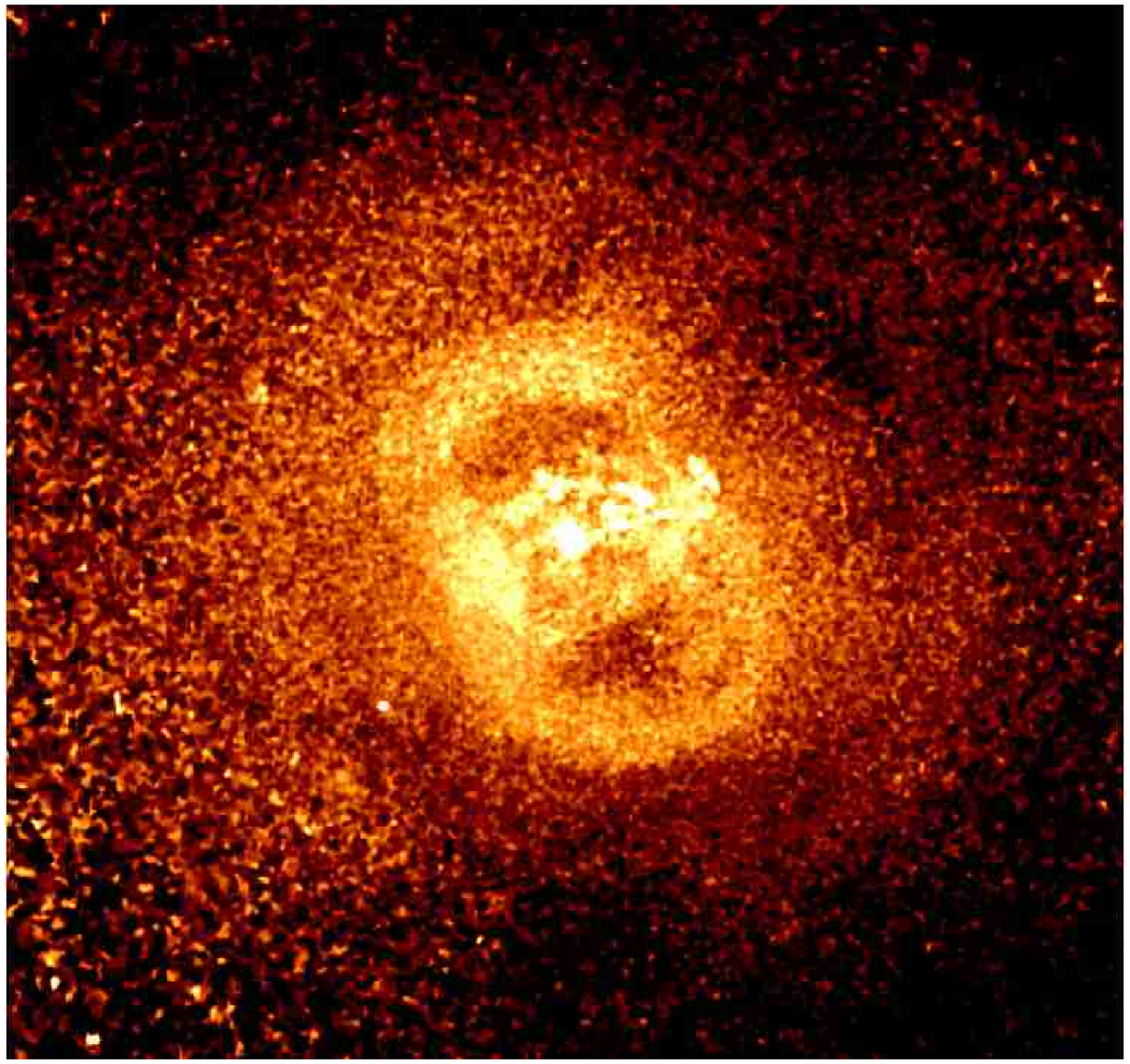}
 \includegraphics[width=0.32\columnwidth]{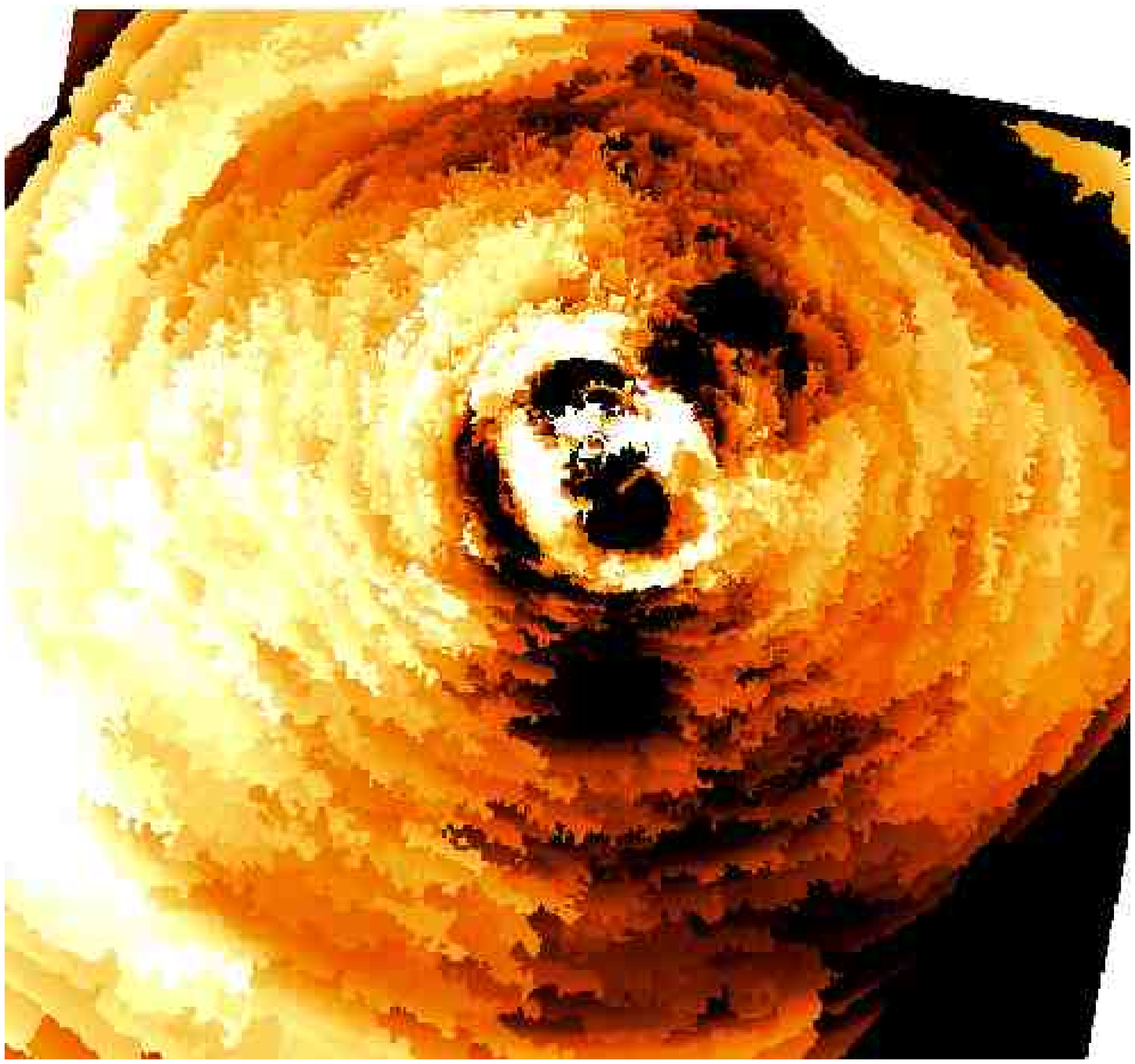}
  \caption{Perseus pressure maps. Left: with radio overlay, Centre:
    showing that the high pressure regions are thick and circular
    around the radio bubbles, Right: pressure difference map showing a
    series of 
    outer bubbles and channels to S and N.}
\end{figure}
\begin{figure}
  \centering
  \includegraphics[width=0.9\columnwidth]{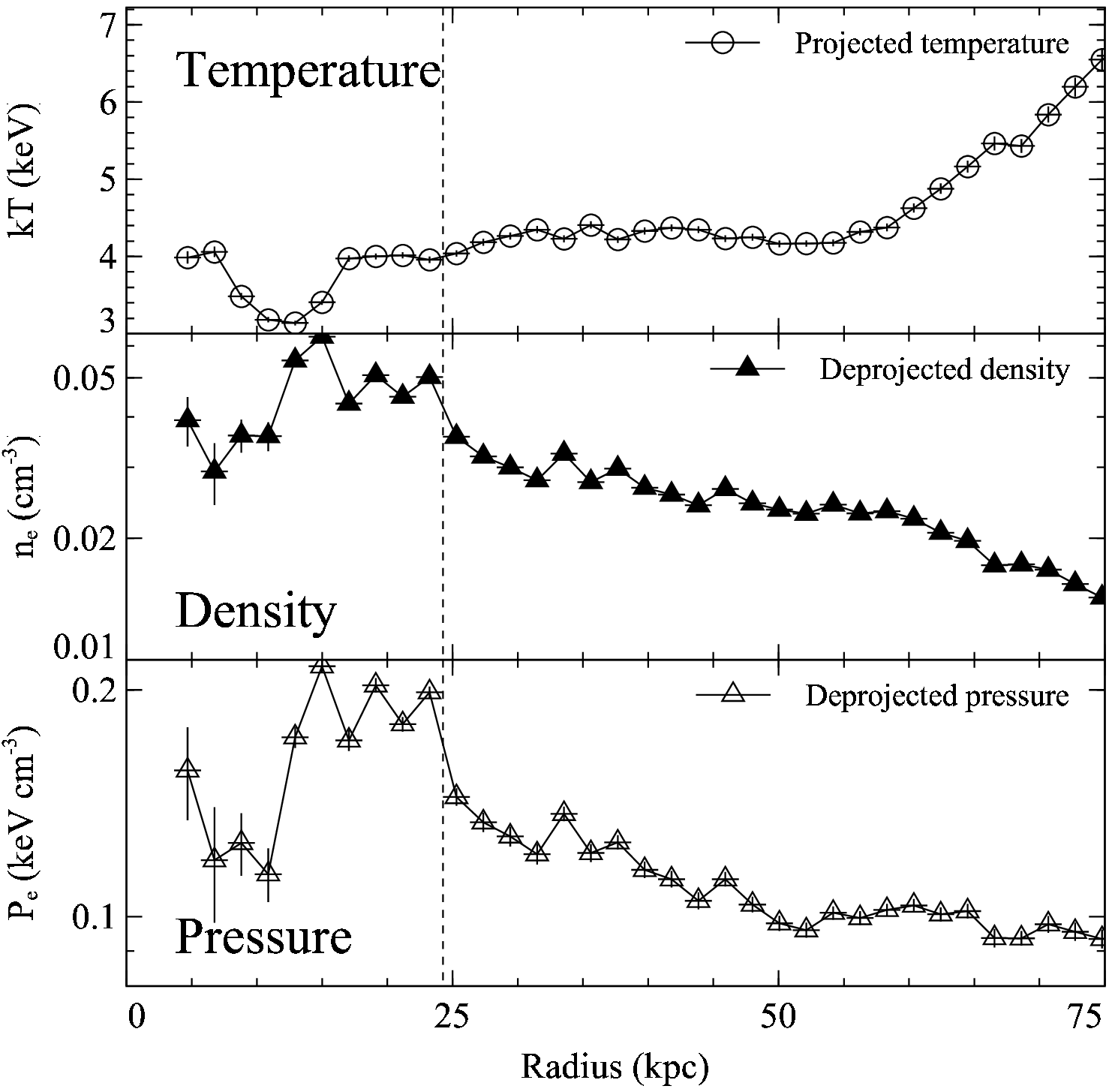}
  \caption{Temperature, density and pressure profiles in a 30 deg wide sector to the
NE of the Northern bubble. The position of the edge of the higher
pressure ring (the shock front) is marked by a
dashed line.}
\end{figure}

With 200~ks Chandra exposure we discovered both cool gas and shocks
surrounding the inner bubbles as well as quasi-circular ripples in the
surrounding gas which we interpreted as sound waves generated by the
cyclical bubbling of the central radio source. Related features have
been seen in the Virgo cluster \cite{For05}\cite{For06}. The NW
ghost bubble has a horseshoe-shaped H$\alpha$ filament trailing it
which shows the streamlines in the hot gas. (The velocities along the
filament are reported by \cite{Hat06} and are consistent with the
streamline hypothesis.) On this basis we concluded that the ICM is not
highly turbulent and thus that viscosity (and possibly conduction) is
high enough to dissipate the energy carried by the sound waves
\cite{Fab03a}\cite{Fab03b}.  Such an energy transport and
dissipation mechanism is roughly isotropic and can thereby provide the
required gently distributed heat source
\cite{Rus04a}\cite{Rus04b}\cite{Rus05}\cite{Rey05}\cite{Fab05}.
The energy flux in the sound waves seen equals the energy required to
balance cooling within a radius of about 100~kpc.

A ripple is presumably made each time a new bubble grows. Close to the
bubble the amplitude must be high so a shock is likely to form, which
could dissipate so much of the energy that there is little left for
significant sound waves and would overheat the innermost
regions \cite{Fuj05}\cite{Mat06}.  We now have 900~ks of good
Chandra data and can see the details of what is happening here. The
ripples (Fig.~6) have an amplitude of $\pm 5$ per cent or less (a
level probably not detectable in any other cluster). Pressure maps
show a thick high pressure rim to the bubbles (Fig.~6) with a sharp
outer edge (best seen to the NE in Fig.~5). We measure the density
jump at this edge as 40\%. If the gas has an adiabatic index of 5/3
this should lead to a temperature rise from 4 to 5~keV. No such
temperature jump is seen. Indeed the gas temperature, if anything,
drops slightly across the edge \cite{Fab06}. (The sharp edge appears all round the
bubbles so it is not a cold front.)

Observationally it seems that the violent pumping action of bubble
growth does {\it not} lead to excessive heating of the innermost
regions and allows sound waves to propagate outward, where the energy
can be dissipated by viscosity and conduction in a more distributed
manner, as required. The lack of a temperature jump is puzzling, but
can be understood if energy is spread by thermal conduction
\cite{Fab06}.

\section{Multiphase effects and cooling}

\begin{figure}
  \centering
  \includegraphics[width=0.9\columnwidth]{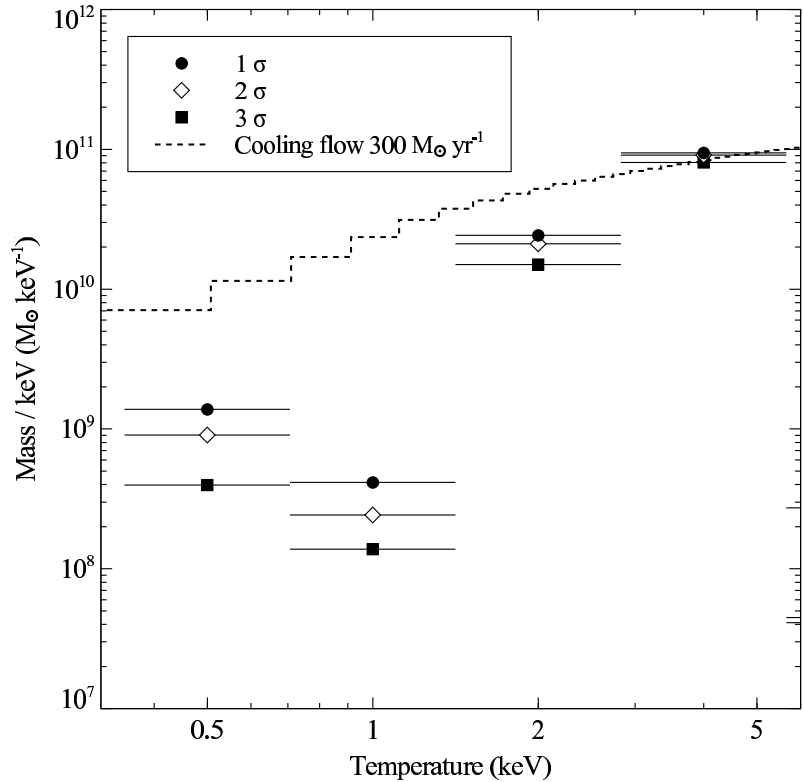}
  \caption{Gas mass as a function of temperature within the central
    1.5 arcmin radius of the Perseus cluster \cite{Fab06}. The expectation for a
    constant pressure cooling flow is shown for comparison.}
\end{figure}

The central galaxy in the Perseus cluster, NGC\,1275, is surrounded by
a spectacular filamentary optical  nebulosity (see \cite{Con01}). This is 
detected in the emission from molecular hydrogen \cite{Hat05} and CO
\cite{Sal05} implying temperatures ranging from several 100 to several
1000~K. Soft X-ray emission is seen coincident with the filaments
requiring a significant mass of gas at ~0.7~keV (Fig.~9). There is
also OVI emission detected by FUSE indicating gas at $\sim5\times
10^5\K$ \cite{Bre06}. Assuming that this last emission is typical and
attributing it all to cooling,  then the cooling rate
in the central 6~kpc is $\sim30\Msunpyr$ \cite{Bre06}. The cooling
rate implied at 1 keV by Fig.~9 is also only about $10\Msunpyr$. Some
mixing, conduction etc could affect these estimates but it would be
difficult to argue for any simple cooling flow over the central region
exceeding about 10\% of that deduced by the hotter X-ray emitting gas,
in the absence of heating. The heating / cooling balance in the
Perseus cluster core, as is the case in most such objects, must be
tightly controlled. 

\section{Discussion}

Heating by the repetitive formation and growth of bubbles has the
quasi-continuous form required to balance radiative cooling. The deep
imaging of the Perseus cluster core reveals the sound waves produced
and shows that the inner shock is isothermal. More work is required to
understand this, but electron conduction is a plausible explanation.

A major remaining problem is how the necessary tight feedback is
achieved. Just how does the central accretion rate adjust to make
heating balance cooling over a $\sim100$~kpc radius region?

\section{Acknowledgements}
We thank Robert Dunn and others for collaboration. ACF thanks the
Royal Society for support.



\end{document}